# Low volume fraction of high-$T_c$ superconductivity in La$_3$Ni$_2$O$_7$ at 80 K and ambient pressure

Mengwu Huo,[1] Peiyue Ma,[1] Chaoxin Huang,[1] Xing Huang,[1] Hualei Sun[2], Meng Wang[1*]

[1]Center for Neutron Science and Technology, Guangdong Provincial Key Laboratory of Magnetoelectric Physics and Devices, School of Physics, Sun Yat-Sen University, Guangzhou, Guangdong 510275, China

[2]School of Science, Sun Yat-Sen University, Shenzhen, Guangdong 518107, China

*corresponding author: wangmeng5@mail.sysu.edu.cn

## Abstract

The discovery of superconductivity in pressurized La$_3$Ni$_2$O$_7$ with a transition temperature of approximately 80 K above the boiling point of liquid nitrogen has sparked significant attention. It is essential to search for high-temperature superconductivity in bulk samples and at ambient pressure in nickelates. In this study, we report influential factors that affect the appearance of superconductivity in La$_3$Ni$_2$O$_7$ at ambient pressure. From direct-current magnetic measurements, we observe a clear diamagnetic response at 80 K in post-annealed single crystals of La$_3$Ni$_2$O$_7$ in oxygen. The superconducting volume fraction is estimated to be within 0.2%, resulting in a decrease in resistivity. This work presents a practical approach for further investigating high-temperature superconductivity in nickelates at ambient pressure.

## Introduction

Since the discovery of superconductivity, it has been a focal research topic in condensed matter physics. However, the origin of high-temperature superconductivity remains complex. One strategy for understanding this phenomenon is to explore new superconductors. Recently, a breakthrough was made in the Ruddlesden-Popper (RP) phase bilayer nickelate La$_3$Ni$_2$O$_7$, where superconductivity was observed with a superconducting (SC) transition temperature ($T_c$) up to 80 K under high pressures between 14 and 43.5 GPa[1]. This discovery has sparked intensive research in experimental and theoretical studies of the RP phase nickelates[2]. Further experimental investigations have confirmed the high-$T_c$ superconductivity, such as zero resistance and diamagnetism[3-6]. Later, bulk superconductivity was identified in both single crystals and powder forms of the La$_3$Ni$_2$O$_7$ system[7-11]. The SC volume fraction is up to 97% in Pr-doped La$_2$PrNi$_2$O$_7$ powder samples[7]. Superconductivity was also detected in pressurized trilayer RP phase nickelates La$_4$Ni$_3$O$_{10}$ and Pr$_4$Ni$_3$O$_{10}$ with the maximum $T_c$ up to 30 K under high pressures[10-18]. The RP phase of nickelates differs from the oxygen reduced PR nickelates, which have formulas of $R$NiO$_2$ and $R_6$Ni$_5$O$_{12}$ ($R$ = rare earth metals) with Ni-3$d^9$ electrons [19-24]. Therefore, the RP phase nickelates provide a new platform for exploring the high-$T_c$ superconductivity and its mechanism.

Experimental investigations on La$_3$Ni$_2$O$_7$ at ambient pressure reveal density wave-like transitions at ~153 and 110 K, which were suggested to originate from a spin density wave and charge density wave, respectively[25-28]. Under pressure, they have distinct behaviors, where the former is enhanced, and the latter is suppressed[29]. Their connections to pressure-induced superconductivity are unknown[29, 30]. The strengths of the electronic correlations in the bilayer and trilayer RP phases of nickelates are comparable with those of cuprates, but the magnetic exchange interactions are different. La$_3$Ni$_2$O$_7$ has a strong interlayer coupling of ~60 meV and weak intralayer couplings of ~3-5 meV[27, 31]. The Ni-3$d_{x2-y2}$ orbitals cross the Fermi level, while the Ni-3$d_{z2}$ orbital is below 50 meV of the Fermi level[32]. However, the superconductivity was observed under high pressure. A structural transition accompanies the emergence of superconductivity[13, 33, 34]. The physical properties are difficult to measure under high pressures in the superconducting state, preventing elucidation of the superconductivity mechanism in the RP phase of nickelates. Therefore, searching for superconductivity in the RP phase of nickelates at ambient pressure is crucial.

Occupancy and distortion of the oxygen ions in the NiO$_6$ octahedra are important for the crystal field splitting, electronic correlations, and magnetic exchange interactions[35-43]. In this study, we have identified an effective strategy to induce superconductivity in La$_3$Ni$_2$O$_7$ at ambient pressure by annealing the single crystal samples in high-pressure O$_2$. Through direct-current (d.c.) magnetic susceptibility measurements, we have observed a clear diamagnetic response at approximately 80 K. Additionally, the electrical resistivity shows metallic behavior with an additional decrease from 80 to 20 K, indicating the presence of a SC phase with a low SC volume fraction in La$_3$Ni$_2$O$_7$ at ambient pressure. Our study provides a potential pathway for achieving bulk superconductivity in nickelates under ambient pressure.

## Results

The bilayer La$_3$Ni$_2$O$_7$ crystallizes into the orthorhombic *Amam* space group, with lattice parameters of $a$ = 5.407(2) Å, $b$ = 5.4176(18) Å, and c = 20.490(5) Å. As shown in Fig. 1(a), the double NiO$_6$ octahedral layers with shared apical oxygens are separated by the LaO layer. X-ray diffraction measurements on single crystal samples confirm the bilayer structure of La$_3$Ni$_2$O$_7$, as shown in Figs. 1(b)-(d).

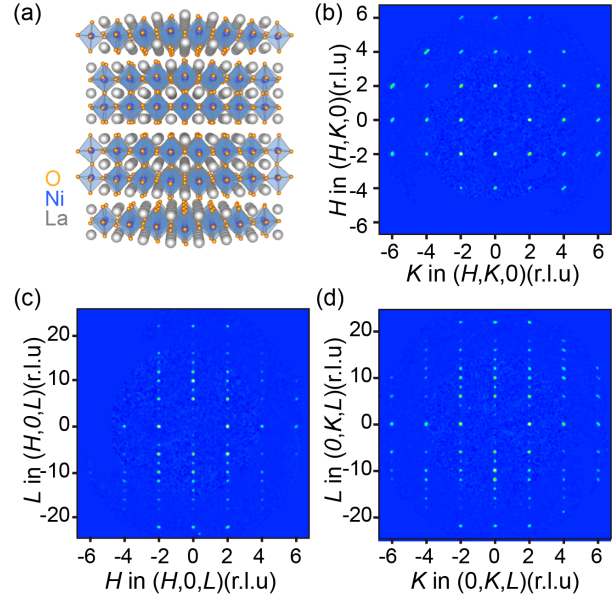

Fig.1 Crystal structure and X-ray reflection peak patterns of La$_3$Ni$_2$O$_7$. (a) Crystal structure of La$_3$Ni$_2$O$_7$ at ambient pressure with an orthorhombic *Amam* space group. (b) Single crystal X-ray diffraction patterns in the (*H, K,* 0) plane, (c) the (*H,* 0, *L*) plane, and (d) the (0, *K, L*) plane of La$_3$Ni$_2$O$_7$.

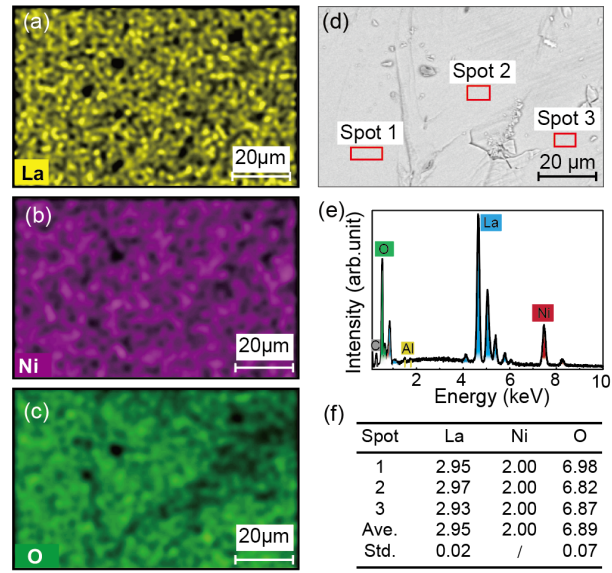

Fig.2 EDS measurements on La$_3$Ni$_2$O$_7$. Distribution of the selected elements of (a) La, (b) Ni, and (c) O in a single crystal of La$_3$Ni$_2$O$_7$. (d) A zoomed-in view on the surface of the La$_3$Ni$_2$O$_7$ sample, captured by SEM. (e) The EDS spectrum of the spot 1 in (d). (f) The compositions measured on the three spots (red squares) in (d) with EDS. The content of Ni has been normalized to 2.

The compositions of the La$_3$Ni$_2$O$_7$ single crystals were analyzed using Energy-dispersive spectroscopy (EDS) in both spot and mapping modes. A scanning electron microscope (SEM) image in Fig. 2(d) provides a detailed view of the surface of the La$_3$Ni$_2$O$_7$. The elemental distribution for the selected elements La, Ni, and O are shown in Figs. 2(a)-2(c), indicating that the distribution of elements is uniform. Figure 2(e) presents the EDS spectrum from spot 1 marked in the SEM image. The elements of carbon and aluminum are from carbon tape and aluminum holders. Figure 2(f) displays the compositional results from the three marked spots on the SEM image in Fig. 2(d). With the content of Ni normalized to 2, the composition determined from EDS is $La_{2.95}Ni_2O_{6.89}$ (Fig. 2f), consistent with the nominal composition of $La_3Ni_2O_7$.

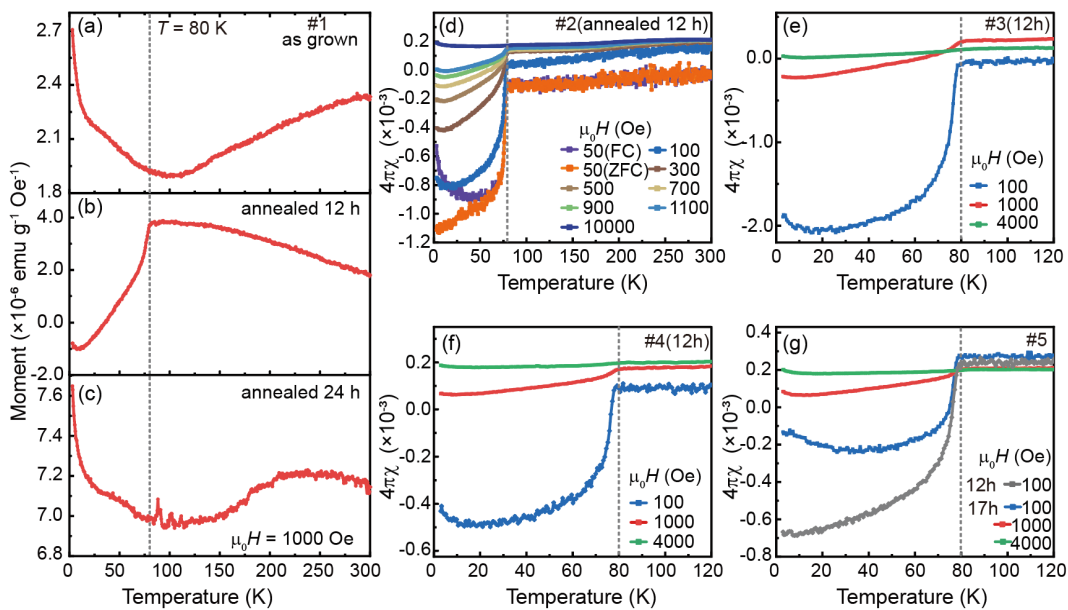

Fig.3 Temperature-dependent d.c. magnetic susceptibility as a function of temperature for $La_3Ni_2O_7$ on as-grown and annealed single crystals. (a) Magnetic susceptibility versus temperature down to 3 K for the as-grown crystal named as sample #1, (b) annealed in a 10 MPa $O_2$ atmosphere and 500 °C for 12 hours, and (c) annealed in the same condition for 24 hours and measured under the zero-field cooled (ZFC) mode with a 1000 Oe field. Normalized temperature-dependent magnetic susceptibility curves under various d.c. magnetic fields below 1 T on (d) sample #2, (e) sample #3, (f) sample #4, and (g) sample #5. The annealing time for the samples has been labeled in the figures. The vertical dashed lines indicate 80 K.

It is known that the oxygen content is intimately connected with the electrical property of $La_3Ni_2O_7$; we thus investigated the annealing effect in $O_2$ gas. An as-grown single crystal sample (#1, 2.8 mg) with a similar magnetic susceptibility [Fig. 3(a)] to those that showed superconductivity under pressures was selected. The sample #1 was annealed in $O_2$ atmosphere at 10 MPa for 12 h. The d.c. zero-field-cooled (ZFC) magnetic susceptibility drops below 80 K. Although a large part of the susceptibility is positive, the onset of the decrease at 80 K is the same as the SC transition temperature $T_c$ determined from resistance measurements under pressure in $La_3Ni_2O_7$. This unexpected signal is reminiscent of the SC transition observed in cuprate superconductors with low SC volume fraction and indicates the emergence of superconductivity at ambient pressure. Sample #1 was annealed in the same condition for another twelve hours to improve the possible SC volume fraction. However, the ZFC susceptibility in Fig. 3(c) yielded no diamagnetic response. The results suggest that the detailed annealing process is critical in inducting superconductivity in $La_3Ni_2O_7$ at ambient pressures.

To confirm the validity of the annealing effect, sample #2, with a mass of 5.7 mg,

was annealed in the same condition for twelve hours. As shown in Fig. 3(d), a sharp transition in ZFC susceptibility appears below 80 K, and the susceptibility is negative, consistent with the Meissner effect of a SC transition. If we estimate the SC volume fraction using the relation $\frac{4\pi\rho M}{\mu_0 H} \times 100\%$, it is only 0.1% with a 50 Oe magnetic field. The magnetic fields up to 1 T can suppress the SC transition without lowering the $T_c$. The diamagnetic signal of superconductivity may be drowned in the paramagnetic background.

As shown in Figs. 3(e)-3(g), we repeated the treatment and measurements on more samples (#3 3.3 mg, #4 6.5 mg, #5 6.8 mg). They all show a clear diamagnetic response with almost the same onset transition temperature of 80 K. To optimize the treatment process, sample #5 was annealed further for five hours. However, the diamagnetic response was weakened, as shown in Fig. 3(g). The largest SC volume fraction was estimated to be 0.2% in sample #3 [Fig. 3(e)]. The transition in the ZFC susceptibility should be attributed to the diamagnetic response of low SC volume fraction in $La_3Ni_2O_7$ at ambient pressure, that is, the Meissner effect of the superconductivity in the annealed samples.

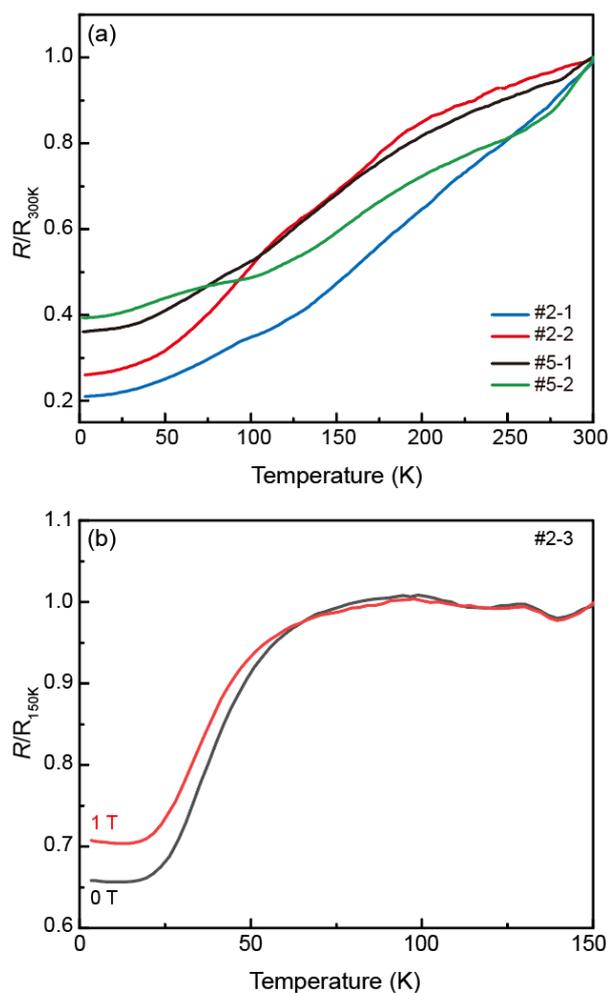

Fig.4 Electric transport properties of the annealed $La_3Ni_2O_7$ single crystals. (a)Temperature dependence of the normalized $R(T)/R_{300 K}$ for annealed $La_3Ni_2O_7$ in the temperature range of 3–300

K. (b) Normalized resistance for sample #2-3 with a magnetic field of $\mu_0H$=0 and 1 T applied in the temperature range of 3–150 K.

It is crucial to demonstrate a SC transition by measuring zero resistivity. The crystals with a diamagnetic response in the d.c. magnetic susceptibility measurements were cut into small pieces for electrical measurements. Samples #2-1, #2-2, and #2-3 were taken from sample #2, which was annealed in a 10 MPa oxygen atmosphere at 500 °C for 12 hours. The samples #5-1 and #5-2 were taken from a slightly larger sample #5, which was annealed for 17 hours. The normalized resistance $R(T)/R_{300K}$ of five pieces of $La_3Ni_2O_7$ single crystals are presented in Fig. 4. They all show metallic behavior as the as-grown samples. Figure 4(b) displays the normalized electrical resistance of sample #2-3. As the temperature decreases, there is a broad drop in resistance (~ 30%) from 80 to 20 K. A plateau appears below 20 K reminiscent of the emergence of the superconductivity in $La_3Ni_2O_7$ under pressure. This decrease in resistance can be attributed to SC islands within the $La_3Ni_2O_7$ crystals. Considering the low volume fraction of superconductivity, zero resistance could not be obtained on these samples.

## Discussion

Combining the magnetic susceptibility and electronic resistance results, it seems convincing that a low volume fraction of superconductivity at 80 K and ambient pressure has been obtained. What is the connection between the SC phases at ambient pressure and high pressure? What are the possible reasons for the appearance of SC at ambient pressure? We conducted scanning transmission electron microscopy (STEM) measurements on sample #2. The main phase is the bilayer structure of $La_3Ni_2O_7$[35]. The $T_c$ is identical for the SC phases at ambient and high pressure; superconductivity from the other RP phases can be excluded. Superconductivity from the interfacial phases between different RP phases is also not likely because intergrowth is common for the RP phases of nickelates[44-47].

The superconductivity should be attributed to the strained bilayer structural phase of $La_3Ni_2O_7$. Recently, Zhou et al. reported nanoscale stripe structures on $La_3Ni_2O_7$ single crystals[48]. The samples are the same as the as-grown samples used in this work. The nanoscale stripe forms micrometer-scale domains. The boundary between the domains can be visualized easily by scattering-type scanning near-field optical microscopy (SNOM) or polarized microscopic spectrometer. Although the microscopic origin of the nanoscale stripe in single crystals of $La_3Ni_2O_7$ is still under investigation, the boundary of the micrometer-scale domains is expected to have a strain effect like the hydrostatic pressure effect on bulk samples. In addition, the synthesis using the high-pressure optical floating zone method can result in residual stress within the single crystals.

Oxygen vacancies can be detrimental to the superconductivity in $La_3Ni_2O_7$. As revealed by STEM, the oxygen vacancies prefer the inner apical oxygen site, which plays a key role in the interlayer superexchange interaction[35]. The inhomogeneity and oxygen vacancies in $La_3Ni_2O_7$ are difficult to avoid and may hinder the superconductivity at ambient and high pressures[7]. Annealing the single crystalline samples in a high-pressure oxygen environment can decrease the oxygen vacancies, which may lead to the emergence of superconductivity in the boundary of the domains. In this scenario, the superconductivity at ambient pressure is filamentary, consistent with the low SC volume fraction and upper critical magnetic field.

However, the reason why the superconductivity is sensitive to the annealing time length is unclear. While preparing this work, we noted that the superconductivity with $T_c$ of 26~42 K was reported in $La_3Ni_2O_7$ thin films[49, 50]. The emergence of superconductivity is sensitive not only to the strain provided by the substrate but also to the detailed annealing conditions. Although the superconductivity in nickelates is crucial for composition and pressure conditions, it is plausible the superconductivity can be realized in bulk and thin film samples at ambient pressure. More efforts are required to improve the SC volume fraction and sample quality.

## Conclusion

In summary, by annealing the as-grown bilayer structural $La_3Ni_2O_7$ single crystals, we have measured the diamagnetic susceptibility below 80 K, which can be attributed to the Meissner effect associated with superconductivity at ambient pressure. The maximum SC volume fraction is estimated to be 0.2%. Although zero resistance was not observed, a decrease in resistance and a plateau at low temperatures are consistent with the emergence of superconductivity with a low-volume fraction. The superconductivity appears filamentary and may originate from the boundary of the domains within the stripe phase. Optimizing the single crystal growth procedure and refining the annealing conditions may enhance the SC volume fraction at ambient pressure.

## Methods

### Crystal growth and characterization

Precursor powders of $La_3Ni_2O_7$ were synthesized through a conventional solid-state reaction using high-purity $La_2O_3$ (99.99%) and NiO (99.9%) as starting materials. The raw materials were thoroughly ground and then annealed at 1200 °C for 24 h. The resultant powders were then hydrostatically pressed into an 8 cm long and 6 mm in diameter rod and sintered for 48 h at 1400 °C. $La_3Ni_2O_7$ single crystals were grown using a floating-zone furnace (HKZ-100, SciDre) at a $p(O_2)$ = 15 bar with a 5 kW Xenon arc lamp. The growth rate was at 3~5 mm/h with a flow rate of 0.1~0.2 L/min of oxygen. The single crystals used for measurements were oxidized in an atmosphere of $O_2$ at a pressure of approximately 10~11 MPa and a temperature of 500 °C.

Single-crystal X-ray diffraction patterns were collected using the Cu-$K_\alpha$ radiation at a temperature of 80 K using a single-crystal X-ray diffractometer (SuperNova, Rigaku). The temperature was controlled by flowing nitrogen gas.

Magnetic susceptibility measurements were performed on the single crystals using a physical property measurement system (PPMS, Quantum Design). The specimens were attached to a quartz holder using a small amount of glue. The ZFC magnetization was recorded on warming at a 3 K/min rate under a fixed field.

The resistivity was measured using a four-probe method, with contacts established by attaching gold thread with conductive silver paint. The sample used for resistance measurements was a smaller piece cut from the larger sample that exhibits a diamagnetic signal.

## Acknowledgements

This work was supported by the National Natural Science Foundation of China (Grants No. 12494591, No. 12425404, No. 12474137, and No. 12174454), the National Key Research and Development Program of China (Grants No. 2023YFA1406000, 2023YFA1406500), the Guangdong Basic and Applied Basic Research Funds (Grants No. 2024B1515020040, 2024A1515030030), Shenzhen Science and Technology Program (Grant No. RCYX20231211090245050), Guangzhou Basic and Applied Basic

Research Funds (Grant No. 2024A04J6417), Guangdong Provincial Key Laboratory of Magnetoelectric Physics and Devices (Grant No. 2022B1212010008), and Research Center for Magnetoelectric Physics of Guangdong Province (2024B0303390001).

**Author contributions**

M. W. supervised the research. M. H. grew and annealed the single crystal samples and performed the resistance and susceptibility measurements with the support of P. M., X. H., C. H., and H.S.. M. W and M. H. wrote the paper with inputs from all authors.